\documentclass[onecolumn,preprintnumbers,amsmath,amssymb,prb]{revtex4}

\usepackage{graphicx}
\usepackage{dcolumn}
\usepackage{bm}
\usepackage{amssymb}
\usepackage{epstopdf}
\usepackage{color}
\usepackage{dsfont}
\usepackage{float}

\begin{document}

\title{Structure of surface electronic states in strained mercury telluride}

\author{O. V. Kibis$^{1,2}$}\email{Oleg.Kibis(c)nstu.ru}
\author{O. Kyriienko$^{3,4}$}
\author{I. A. Shelykh$^{2,4}$}
\affiliation{${^1}$Department of Applied and Theoretical Physics,
Novosibirsk State Technical University, Karl Marx Avenue 20,
Novosibirsk 630073, Russia} \affiliation{$^2$ Science Institute,
University of Iceland IS-107, Reykjavik, Iceland}
\affiliation{${^3}$NORDITA, KTH Royal Institute of Technology and
Stockholm University, Roslagstullsbacken 23, SE-106 91 Stockholm,
Sweden}\affiliation{$^4$ITMO University, Saint Petersburg 197101,
Russia}
%\date{\today}

\begin{abstract}
We present the theory describing the various surface electronic
states arisen from the mixing of conduction and valence bands in a
strained mercury telluride (HgTe) bulk material. We demonstrate
that the strain-induced band gap in the Brillouin zone center of
HgTe results in the surface states of two different kinds. Surface
states of the first kind exist in the small region of electron
wave vectors near the center of the Brillouin zone and have the
Dirac linear electron dispersion characteristic for topological
states. The surface states of the second kind exist only far from
the center of the Brillouin zone and have the parabolic dispersion
for large wave vectors. The structure of these surface electronic
states is studied both analytically and numerically in the broad
range of their parameters, aiming to develop its systematic
understanding for the relevant model Hamiltonian. The results
bring attention to the rich surface physics relevant for
topological systems.
\end{abstract}

\maketitle

\section{Introduction}
The studies of two-dimensional (2D) electronic modes localized
near the surface of a three-dimensional (3D) condensed matter
structure (surface electronic states) represent one of the most
actively studied directions of modern science of the last decade.
The increasing interest of scientific community to this research
area is caused by the topologically nontrivial nature of the
surface states in structures of certain type known as topological
insulators (TIs). Namely, TIs are condensed matter systems which
behave like an insulator in their 3D bulk but have 2D gapless
conducting electronic states protected by the time-reversal
symmetry at their boundaries~\cite{Hasan_2010}. Up to date, such
topologically protected electronic states were intensively studied
theoretically and experimentally in various condensed matter
structures~\cite{Kane_2005,Fu_2007,Fu_2007_1,Moore_2007,Roy_2009,Xia_2009,Zhang_2009,Chen_2009,Yudin_2016,Hasan_2017,Pervishko_2018,Kozin_2018},
and their optical analogs were also
revealed~\cite{Lu2014,Mittal2016,Whittaker2018,Klembt2018,Banerjee2018}.

Particularly, it follows from the theoretical analysis based on
the $Z_2$ topological invariants that the surface electronic
states in bulk mercury telluride (HgTe) can be topologically
nontrivial~\cite{Fu_2007}. However, the band structure of natural
HgTe is semi-metallic: There is the small overlap of conduction
and valence bands originated from the bulk inversion asymmetry of
the crystal structure~\cite{Bir_Pikus}. In order to observe the
predicted topological surface states in HgTe, one needs to turn
this semi-metal into insulator. To solve the problem, an uniaxial
strain as a tool to open the band gap between the valence and
conduction bands of HgTe was proposed~\cite{Dai_2008}. Following
this methodology, the topological surface states in strained HgTe
were recently observed experimentally within the strain-induced
band gap~\cite{Brune_2011}. As a consequence, physical properties
of the surface electronic states in strained HgTe-based materials
are currently in the focus of
attention~\cite{Maier_2012,Yan_2012,Chen_2012,Kozlov_2014,Dantscher_2015,Li_2015,Ruan_2016,Kirtschig_2016,Leubner_2016,Haas_2017,Grendysa_2017,Thomas_2017}.
However, the theory describing the structure of these surface
states is still far from being complete. Particularly, the
characteristic feature of HgTe is the coexistence of surface
electronic states of different physical nature: Besides of the
discussed topological surface states in gapped HgTe, there are the
surface states in gapless HgTe analyzed for the first time by
D'yakonov and Khaetskii~\cite{Dyakonov_1981}. Certainly, the
consistent theory should be able to describe the dependence of all
surface states on the strain. The present article takes a step
towards such a consistent theory and provides an intuitive
understanding of the system.

The paper is organized as follows. In Sec. II, we formulate a
formalism describing the surface electronic states of various
kinds in strained HgTe. In Sec. III, we solve the corresponding
Schr\"odinger problem analytically in the simplest particular
cases, calculate the dispersion of the surface states numerically,
and analyze their energy spectrum. The last two sections contain
the conclusion and acknowledgments.

\section{Model}
We consider the surface electronic states which originate from the
mixing of conduction and valence bands of HgTe near the center of
the Brillouin zone (the electronic term $\Gamma_8$). In bulk
strained HgTe, the states of this electronic term are described by
the Hamiltonian~\cite{Bir_Pikus}
\begin{equation}\label{H}
\hat{\cal H}=\hat{\cal H}_{\mathrm{L}}+\hat{\cal
H}_{\mathrm{strain}}+\hat{\cal H}_{\mathrm{BIA}}.
\end{equation}
where
\begin{equation}\label{HL}
\hat{\cal
H}_{\mathrm{L}}=\left(\gamma_1+\frac{5}{2}\gamma_2\right)
\mathbf{k}^2-2\gamma_2(J_x^2k_x^2+J_y^2k_y^2+J_z^2k_z^2)-2\gamma_3(\{J_x,
J_y\}k_x k_y+ \{J_x, J_z\}k_x k_z+\{J_y, J_z\}k_y k_z)
\end{equation}
is the conventional Luttinger Hamiltonian describing the
conduction and valence bands of unstrained HgTe,
\begin{equation}\label{Hstr}
\hat{\cal H}_{\mathrm{strain}}=\left(a+\frac{5}{4}b\right)
(u_{xx}+u_{yy}+u_{zz})-b(J_x^2u_{xx}+J_y^2u_{yy}
+J_z^2u_{zz})-\frac{d}{\sqrt{3}}(\{J_x, J_y\}u_{xy}+ \{J_x,
J_z\}u_{xz}+\{J_y, J_z\}u_{yz})
\end{equation}
is the Bir-Pikus Hamiltonian describing the modification of
conduction and valence bands of HgTe under strain, and
\begin{equation}\label{Hbia}
\hat{\cal H}_{\mathrm{BIA}}=\alpha[k_x\{J_x,
(J_y^2-J_z^2)\}+k_y\{J_y, (J_z^2-J_x^2)\} +k_z\{J_z,
(J_x^2-J_y^2)\}]
\end{equation}
is the term arisen from the bulk inversion asymmetry (BIA) of HgTe
crystal. Correspondingly, $\mathbf{k}=(k_x,k_y,k_z)$ is the
electron wave vector, $\gamma_{1,2,3}$ are the Luttinger
parameters of HgTe, $a$, $b$ and $d$ are the deformation
potentials of HgTe, $u_{ij}$ are the components of the deformation
tensor of the strained HgTe, $\alpha$ is the BIA parameter,
$J_{x,y,z}$ are the $4\times4$ matrices corresponding to the
electron angular momentum $J=3/2$, and the curly brackets
$\{A,B\}$ represent anti-commutators of the matrices  $A$ and $B$.
The eigenstates of the Hamiltonian (\ref{H}) can be written in the
most general form as four independent spinors,
\begin{equation}\label{s}
\varphi_{m}=\left[ C_{1m},\,C_{2m},\,C_{3m}\, C_{4m}\right]^{T}
e^{i\mathbf{k}\mathbf{r}},
\end{equation}
with indices $m=1,2,3,4$, and the spinor components
$C_{nm}(k_x,k_y,k_z)$ are the functions of the electron wave
vector, $\mathbf{k}=(k_x,k_y,k_z)$.

For definiteness, let us consider electronic states localized near
the $(001)$-surface, assuming that the bulk HgTe fills the
half-space for $z>0$. It follows from the conservation laws that
the surface $(001)$ mixes different electronic states of the
Hamiltonian (\ref{H}) with the same energy,
$\varepsilon(\mathbf{k}_s,k_z)$, and the same wave vector in the
surface plane, $\mathbf{k}_s=(k_x,k_y)$, but with different normal
components of the wave vector, $k_z$. As a consequence, the
surface-localized electronic states which arise from the mixing
are described by the same spinors (\ref{s}) with the imaginary
$z$-component of electron wave vector, $k_z=i\kappa$, where
$\kappa>0$ is the localization parameters, corresponding to the
inverse of localization length for electronic states near the
surface. Generally, the localization parameter, $\kappa$, can be a
complex number but its real part must be positive. In bulk HgTe,
there are the four different branches of electron energy,
$\varepsilon(\mathbf{k})$, corresponding to the four branches of
the conduction and valence bands spin-split due to the BIA terms
of the Hamiltonian (\ref{H}). Therefore, there are four different
parameters, $\kappa_{1,2,3,4}(\mathbf{k}_s,\varepsilon)$, which
can be found as four solutions of the secular equation,
\begin{equation}\label{det}
\mathrm{det}\vert \hat{\cal H}(\mathbf{k}_s,i\kappa)-\varepsilon
\vert =0.
\end{equation}
Making the replacement, $k_z\rightarrow i\kappa_{1,2,3,4}$, in the
eigenspinors (\ref{s}), we can write the surface-localized
eigenfunction of the Hamiltonian (\ref{H}) as a linear combination
of the spinors,
\begin{equation}\label{psi}
\Psi=e^{i\mathbf{k}_s\mathbf{r}_s}\sum_{m=1}^{4}\left[
C_{1m}(\mathbf{k}_s,i\kappa_m),C_{2m}(\mathbf{k}_s,i\kappa_m),
C_{3n}(\mathbf{k}_s,i\kappa_m),C_{4n}(\mathbf{k}_s,i\kappa_m)\right]^T
A_me^{-\kappa_mz},
\end{equation}
where $\mathbf{r}_s=(x,y)$ is the electron radius-vector in the
surface plane and $A_{1,2,3,4}$ are the constants to be
determined. To do so, we chose the model of a surface potential
which can be approximated by the infinitely-high barrier at
position $z=0$. This sets the boundary condition for the electron
wave function (\ref{psi}) as $\Psi|_{z=0}=0$, and results into the
homogeneous system of four algebraic equations defining the
constants $A_j$,
\begin{equation}\label{eq}
\sum_{m=1}^{4}A_mC_{nm}(\mathbf{k}_s,i\kappa_m)=0,\quad
(n=1,2,3,4).
\end{equation}
The secular equation of the algebraic system (\ref{eq}),
\begin{equation}\label{dett}
\mathrm{det}\vert C_{nm}(\mathbf{k}_s,i\kappa_m)\vert =0, \quad
(n,m=1,2,3,4),
\end{equation}
defines the sought energy spectrum of the surface electronic
states, $\varepsilon(\mathbf{k}_s)$. In the next section of the
article, we apply the strategy described above to find the
spectrum $\varepsilon(\mathbf{k}_s)$ in systems with various band
parameters. In the case of HgTe, we use the following
parameters~\cite{Adachi_2004,Ruan_2016}:
$\gamma_1=15.6\,\hbar^2/2m_0$, $\gamma_2=9.6\,\hbar^2/2m_0$,
$\gamma_3=8.6\,\hbar^2/2m_0$, $b=-1.22$~eV and
$\alpha=0.208$~\AA$\cdot$eV.

\section{Results and Discussion}

In what follows, we consider the Hamiltonian (\ref{H}) to be
written as a $4\times4$ matrix in the conventional Luttinger-Kohn
basis, $\psi_{j_z}$, corresponding to the different projections of
electron angular momentum on the $z$ axis, $j_z=\pm1/2,\pm3/2$
(see the Appendix for details). In unstrained bulk HgTe, the four
wave functions, $\psi_{j_z}$, are the eigenfunctions of the
Hamiltonian (\ref{H}): They describe the states of the electronic
term $\Gamma_8$ at $\mathbf{k}=0$, which are four-fold degenerate.
As for strained bulk HgTe, we consider for definiteness the case
of an uniaxial stress applied along the $z$ axis. Under the
uniaxial stress, the deformation tensor of HgTe, $u_{ij}$, written
in the principal crystallographic axes, $x,\,y,\,z$, is diagonal.
Particularly, the case of $u_{zz}<0$ and $u_{xx}=u_{yy}>0$
corresponds to the compressive strain, whereas the opposite case
of $u_{zz}>0$ and $u_{xx}=u_{yy}<0$ corresponds to the tensile
strain. As a result, the strain Hamiltonian (\ref{Hstr}) can be
rewritten as
\begin{equation}\label{str}
\hat{\cal
H}_{\mathrm{strain}}=a(u_{xx}+u_{yy}+u_{zz})+(\varepsilon_g/2)(J_z^2-5/4),
\end{equation}
where $\varepsilon_g=2b(u_{xx}-u_{zz})=2b(u_{yy}-u_{zz})$, and we
note that $b<0$ for HgTe. The first term of the Hamiltonian
(\ref{str}) is the strain-induced shift of zero energy, which will
be omitted in the following. As for the second term, it describes
the strain-induced splitting of the electronic states with
$j_z=\pm1/2$ and $j_z=\pm3/2$ at $\mathbf{k}=0$. It follows from
the total Hamiltonian (\ref{H}) with the strain Hamiltonian
(\ref{str}) that
$\varepsilon_g=\varepsilon_{\pm3/2}-\varepsilon_{\pm1/2}$, where
$\varepsilon_{j_z}$ are the energies of these states at
$\mathbf{k}=0$ (see Fig.~\ref{Fig1}a). It should be noted,
particularly, that the compressive strain ($\varepsilon_g<0$) and
the tensile strain ($\varepsilon_g>0$) result in the opposite
sequence orders for energies of the basic states with the wave
functions $\psi_{\pm1/2}$ and $\psi_{\pm3/2}$. Substituting the
strain Hamiltonian ({\ref{Hstr}) into the total Hamiltonian
(\ref{H}), we can apply the methodology developed in Sec.~II to
analyze the evolution of the surface electronic states in HgTe
under stress.
%%%
\begin{figure}[h]
\includegraphics[width=1\linewidth]{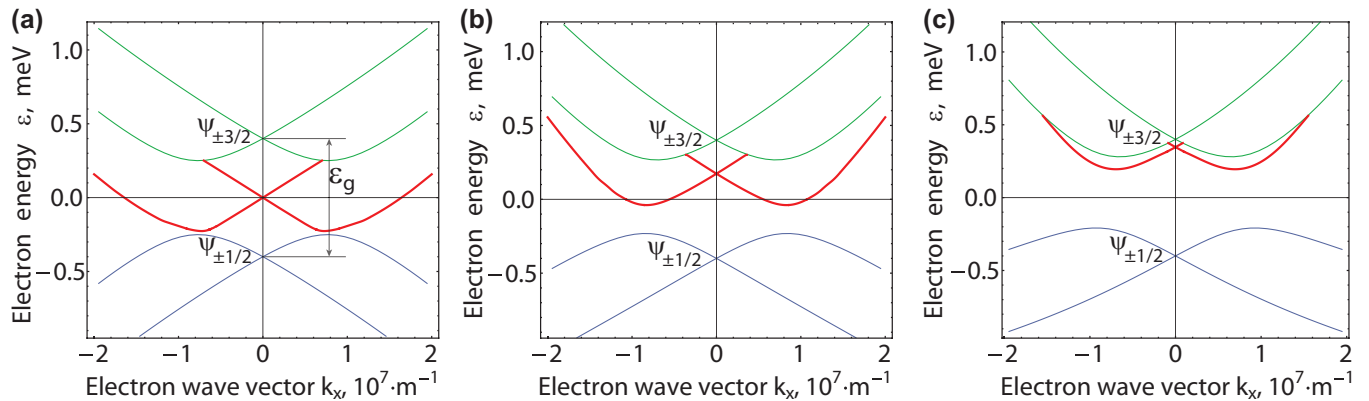}
\caption{Structure of the surface electronic states (bold red
curves) near the Brillouin zone center for the strain-induced gap
$\varepsilon_g= 0.8$~meV and the different values of the Luttinger
parameter $\gamma_1$: (a) $\gamma_1=0$; (b)
$\gamma_1=7.8\,\hbar^2/2m_0$; (c) $\gamma_1=15.6\,\hbar^2/2m_0$
(the case of HgTe). Thin green and blue curves represent the
dispersion of bulk conduction and valence bands which originate
from the terms $\psi_{\pm3/2}$ and $\psi_{\pm1/2}$, respectively.}
\label{Fig1}
\end{figure}
%%%

First of all, let us consider the electronic states with the zero
planar electron wave vector, $\mathbf{k}_s=0$, which are localized
near the $(001)$-surface of the uniaxially strained HgTe. The
simplest case, which can be readily solved, corresponds to the
Luttinger parameter $\gamma_1=0$. Physically, this model situation
describes a semiconductor with the Hamiltonian (\ref{H}), where
the masses of electrons and holes along the $z$ axis, $m_e$ and
$m_h$, are equal to each other,
$m_e/m_h=(2\gamma_2-\gamma_1)/(2\gamma_2+\gamma_1)$. For such a
symmetric electron-hole system, the Hamiltonian (\ref{H}) at
$\mathbf{k}_s=0$ can be written in the Luttinger-Kohn basis as a
block-diagonal matrix (see the Appendix),
\begin{equation}\label{Hsur}
\hat{\cal H}_0=
\begin{bmatrix}
\,\,-2\gamma_2\hat{k}_z^2+\varepsilon_g/2\,\, &
\,\,\sqrt{3}\alpha\hat{k}_z\,\,
&\,\,0\,\,&\,\,0\,\,\\
\,\,\sqrt{3}\alpha\hat{k}_z\,\,&\,\,2\gamma_2\hat{k}_z^2-\varepsilon_g/2\,\,&\,\,0\,\,&
\,\,0\,\,\\
\,\,0\,\,&\,\,0\,\,&\,\,2\gamma_2\hat{k}_z^2-\varepsilon_g/2\,\,&
\,\,-\sqrt{3}\alpha\hat{k}_z\,\,\\
\,\,0\,\,&\,\,0\,\,&\,\,-\sqrt{3}\alpha\hat{k}_z\,\,&\,\,-2\gamma_2\hat{k}_z^2+\varepsilon_g/2\,\,
\end{bmatrix},
\,\,
\end{equation}
where $\hat{k}_z=-i\partial/\partial z$ is the operator of the
electron momentum normal to the considered surface. In the
particular case of the Hamiltonian (\ref{Hsur}), the
surface-localized eigenspinors (\ref{psi}) correspond to the
eigenenergy $\varepsilon_0=0$ and can be written as
\begin{eqnarray}\label{Psi}
\Psi_1&=&D(e^{-\kappa_+z}-e^{-\kappa_-z})\,[
1,\,i,\,0,\,0]^T\nonumber\\
\Psi_2&=&D(e^{-\kappa_+z}-e^{-\kappa_-z})\,[ 0,\,0,\,-i,\,1]^T,
\end{eqnarray}
where
\begin{equation}\label{kappa}
\kappa_\pm=\frac{\sqrt{3}\alpha}{4\gamma_2}\pm\sqrt{\left(\frac{\sqrt{3}\alpha}{4\gamma_2}\right)^2-\frac{\varepsilon_g}{4\gamma_2}}
\end{equation}
are the localization parameters of the surface state, and
$D=\sqrt{\sqrt{3}\alpha\varepsilon_g/(6\alpha^2-8\gamma_2\varepsilon_g)}$
is the normalization constant. The eigenspinors (\ref{Psi}) can be
easily verified by direct substitution into the Schr\"odinger
equation, $\hat{\cal H}_0\Psi_{1,2}=\varepsilon_0\Psi_{1,2}$, with
the Hamiltonian (\ref{Hsur}) and the eigenenergy
$\varepsilon_0=0$. It follows from Eq.~(\ref{kappa}) that
$\kappa_-<0$ if $\varepsilon_g<0$. Since the real part of both
localization parameters, $\kappa_\pm$, must be positive, the
surface states (\ref{Psi}) exist only for $\varepsilon_g>0$
(tensile strain). Physically, this means that the existence of the
surface states (\ref{Psi}), first, arises from the BIA terms
($\alpha\neq0$) and, second, it strongly depends on the sequence
order of their parent bulk states, $\psi_{\pm1/2}$ and
$\psi_{\pm3/2}$.

To find the dispersion of the surface states (\ref{Psi}) for small
wave vectors $\mathbf{k}_s$, we have to project the total
Hamiltonian (\ref{H}) to the subspace spanned by these two states,
$\{\Psi_1,\Psi_2\}$. Keeping the terms linear in $\mathbf{k}_s$,
we arrive at the effective Hamiltonian,
\begin{equation}\label{Heff}
\hat{\cal
H}_{\mathrm{eff}}=\varepsilon_0-\frac{3\alpha}{2}(\sigma_xk_x+\sigma_yk_y)-\frac{\sqrt{3}\alpha}{2}
(\sigma_xk_y+\sigma_yk_x),
\end{equation}
where $\sigma_{x,y}$ are the Pauli matrices written in the basis
$\{\Psi_1,\Psi_2\}$. Diagonalizing the Hamiltonian (\ref{Heff}),
we can write the sought energy spectrum of the surface states
(\ref{Psi}) near $\mathbf{k}_s=0$ as
\begin{equation}\label{ED}
\varepsilon(\mathbf{k}_s)=\varepsilon_0\pm\sqrt{3}\alpha\sqrt{k_x^2+k_y^2+\sqrt{3}k_xk_y}.
\end{equation}
It follows from Eq.~(\ref{ED}) that the two degenerate states
(\ref{Psi}) form the Dirac point at $\mathbf{k}_s=0$ with the
energy $\varepsilon_0=0$ and anisotropic linear dispersion near
the point. Applying the methodology of Sec.~II, one can calculate
numerically the dispersion of the surface states in the broad
range of the wave vectors, $\mathbf{k}_s$. To demonstrate the
properties of these surface states in more details, we plotted
their dispersion for the HgTe band parameters $\gamma_{2,3}$ and
$\alpha$ but for different Luttinger parameters $\gamma_1$ (see
Fig.~\ref{Fig1}). In the model case of symmetric electron-hole
system discussed above ($\gamma_1=0$), the Dirac point energy,
$\varepsilon_0=0$, lies exactly in the middle of the bulk states
$\psi_{\pm1/2}$ and $\psi_{\pm3/2}$ (see Fig.~\ref{Fig1}a),
whereas the nonzero Luttinger parameter $\gamma_1$ shifts the
Dirac point energy, $\varepsilon_0$, from the middle towards the
bulk term $\psi_{\pm3/2}$ (see Fig.~\ref{Fig1}b). As a result, the
branches of the surface states are localized near the bulk term
$\psi_{\pm3/2}$ in the real case of HgTe (see Fig.~\ref{Fig1}c),
where the electron-hole system is strongly asymmetric
($m_e/m_h\ll1$). It should be stressed that the effective
Hamiltonian (\ref{Heff}) and the dispersion (\ref{ED}), which were
derived formally for the particular case of $\gamma_1=0$, are
applicable to describe the energy spectrum of surface states near
the Dirac point for any band parameters and the strain-induced
band gap. Namely, the energy of the Dirac point $\varepsilon_0$ is
proportional to the band gap value,
$\varepsilon_0\propto\varepsilon_g$, where the proportionality
constant depends on the bulk band parameters, $\gamma_{1,2,3}$ and
$\alpha$, and turns into zero if $\gamma_1=0$. Taking this into
account, both the analytical expression for the Dirac dispersion
(\ref{ED}) and the numerically calculated dispersion in
Fig.~\ref{Fig1}c can be used to analyze the surface states in
strained HgTe for any gap, $\varepsilon_g>0$. Particularly, the
Dirac velocity, $v_D=\sqrt{3}\alpha/\hbar$, which can be extracted
from the dispersion (\ref{ED}), does not depend on the gap.
%%%
\begin{figure}[h]
\includegraphics[width=1\linewidth]{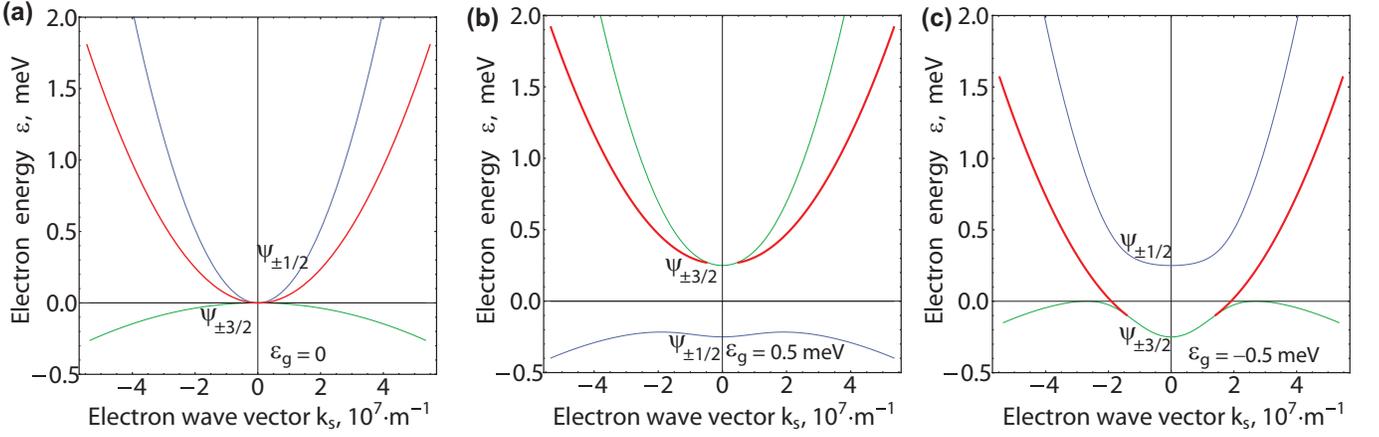}
\caption{Structure of the D'yakonov-Khaetskii surface electronic
states (bold red curves) in HgTe for the different strain-induced
gaps $\varepsilon_g$:(a) unstrained material ($\varepsilon_g=0$);
(b) tensile strain ($\varepsilon_g=0.5$~meV); (c) compressive
strain ($\varepsilon_g=-0.5$~meV). Thin green and blue curves
represent the dispersion of bulk bands which originate from the
terms $\psi_{\pm3/2}$ and $\psi_{\pm1/2}$, respectively.}
\label{Fig2}
\end{figure}
%%%

It follows from the aforesaid that the branches of the surface
states merge into the spectrum of bulk electronic states if the
plane electron wave vector, $\mathbf{k}_s$, is large enough (see
Fig.~\ref{Fig1}c). Therefore, they exist only near
$\mathbf{k}_s=0$. However, there are the surface electronic states
of other kind, which can exist far from $\mathbf{k}_s=0$. In
contrast to the states (\ref{Psi}), the BIA Hamiltonian
(\ref{Hbia}) is not crucial for their existence and will be
omitted in the following analysis. To simplify calculations, we
also will neglect the weak anisotropy of electron-hole dispersion
in HgTe. Following Ref.~\onlinecite{Dyakonov_1981}, this neglect
corresponds to the replacement of the Luttinger parameters,
$\gamma_{2,3}\rightarrow\gamma=(2\gamma_2+3\gamma_3)/5$, in the
Hamiltonian (\ref{HL}). Under these assumptions, the secular
equation (\ref{dett}) reads
\begin{equation}\label{Es}
\left[\lambda_-^{(1)}\lambda_+^{(2)}-1\right]\left[\lambda_+^{(1)}\lambda_-^{(2)}-1\right]=0,
\end{equation}
where
\begin{equation}\label{lambda}
\lambda_\pm^{(j)}=\frac{\sqrt{3}\gamma(k_s^2\mp2k_s\kappa_j)}
{\varepsilon-\gamma_1(k_s^2-\kappa_j^2)+(-1)^j[\gamma(k_s^2+2\kappa_j^2)+\varepsilon_g/2]},
\end{equation}
\begin{equation}\label{ka}
\kappa_j=\sqrt{\frac{A+(-1)^j\sqrt{A^2+4BC}}{2B}},
\end{equation}
and
\begin{equation}\nonumber
A=2\gamma_1(\varepsilon-\gamma_1k_s^2)+2\gamma(4\gamma
k_s^2-\varepsilon_g),\,\,
B=(\gamma_1+2\gamma)(2\gamma-\gamma_1),\,\,
C=(\varepsilon-\gamma_1k_s^2)^2-(\gamma
k_s^2+\varepsilon_g/2)^2-3\gamma^2k_s^4.
\end{equation}
Solving the equation (\ref{Es}) together with Eqs.~(\ref{lambda})
and (\ref{ka}), one can find the sought energy spectrum of the
surface states, $\varepsilon({k}_s)$, under the uniaxial strain.
This spectrum is plotted in Fig.~\ref{Fig2} for the different
strain-induced gaps, $\varepsilon_g$. In the absence of the strain
($\varepsilon_g=0$), Eq.~(\ref{Es}) can be solved analytically and
leads to the parabolic branch of the D'yakonov-Khaetskii (DKh)
surface states~\cite{Dyakonov_1981},
\begin{equation}\label{DKh}
\varepsilon(k_s)=
\left[1-\left(\frac{1+\sqrt{3(2\gamma-\gamma_1)/(2\gamma+\gamma_1)}}{2}\right)^2\right]
(\gamma_1+2\gamma){k_s^2},
\end{equation}
which is plotted in Fig.~\ref{Fig2}a. Solving Eq.~(\ref{Es})
numerically for $\varepsilon_g\neq0$, we can plot the spectrum of
the DKh surface states in strained HgTe. In contrast to the
surface states discussed above, the DKh states exist for both
tensile (see Fig.~\ref{Fig2}b) and compressive (see
Fig.~\ref{Fig2}c) strain. However, their structure is crucially
different for these two cases: In the case of tensile strain, the
DKh states lie only within the bulk spectrum of conduction band
(see Fig.~\ref{Fig2}b), whereas the case of compressive strain
corresponds to the DKh states which lie also within the
strain-induced band gap (see Fig.~\ref{Fig2}c). It should be noted
also that the DKh states exist only for large electron wave
vectors, $k_s$, and vanish near $k_s=0$. It follows from
Figs.~\ref{Fig2}b and \ref{Fig2}c that the DKh branch merges into
the continuum of bulk conduction band in the case of tensile
strain (see Fig.~\ref{Fig2}b) and the continuum of bulk valence
band in the case of compressive strain (see Fig.~\ref{Fig2}c) at a
some critical electron wave vector. The value of the critical wave
vector, $k_s=k_0$, is defined by Eqs.~(\ref{Es})-(\ref{ka}), where
the energy of the surface electron states, $\varepsilon$, is equal
to the energy of one of the two bulk electron branches,
\begin{equation}\label{Es0}
\varepsilon=\gamma_1k_s^2\pm\sqrt{(\gamma
k_s^2+\varepsilon_g/2)^2+3\gamma^2k_s^4}.
\end{equation}
Solving Eqs.~(\ref{Es})-(\ref{ka}) with Eq.~(\ref{Es0}), one can
find that $k_0\propto\sqrt{|\varepsilon_g|}$. Thus, the increase
of the strain-induced gap $\varepsilon_g$ shifts the existence
domain of the DKh surface states to the region of large electron
wave vectors, $k_s$. It follows from Eq.~(\ref{Es}) that the
spectrum of the DKh surface states in strained HgTe is parabolic
and described approximately by Eq.~(\ref{DKh}) for large wave
vectors, $k_s$, satisfying the condition $\gamma
k_s^2\gg\varepsilon_g$. It should be noted that the spectrum of
the DKh surface states in strained materials of HgTe-class
crucially depends on the Luttinger parameters. To demonstrate
this, we plotted the dispersion of the DKh surface states in
Fig.~\ref{Fig3} for the HgTe band parameter
$\gamma=9.0\,\hbar^2/2m_0$ but for different Luttinger parameters
$\gamma_1$. In the case of symmetric electron-hole system
($\gamma_1=0$) there are the two branches of the DKh states which
behave as surface electrons and holes (see Fig.~\ref{Fig3}a). The
nonzero Luttinger parameter $\gamma_1$ merges the upper
(electronic) branch into the continuum of bulk electronic states
and changes the curvature of the lower (hole) branch (see
Fig.~\ref{Fig3}b). As for the real case of HgTe parameters, only
the lower branch survives and turns into the surface states of
electronic kind (see Fig.~\ref{Fig3}c).
%%%
\begin{figure}[h]
\includegraphics[width=1\linewidth]{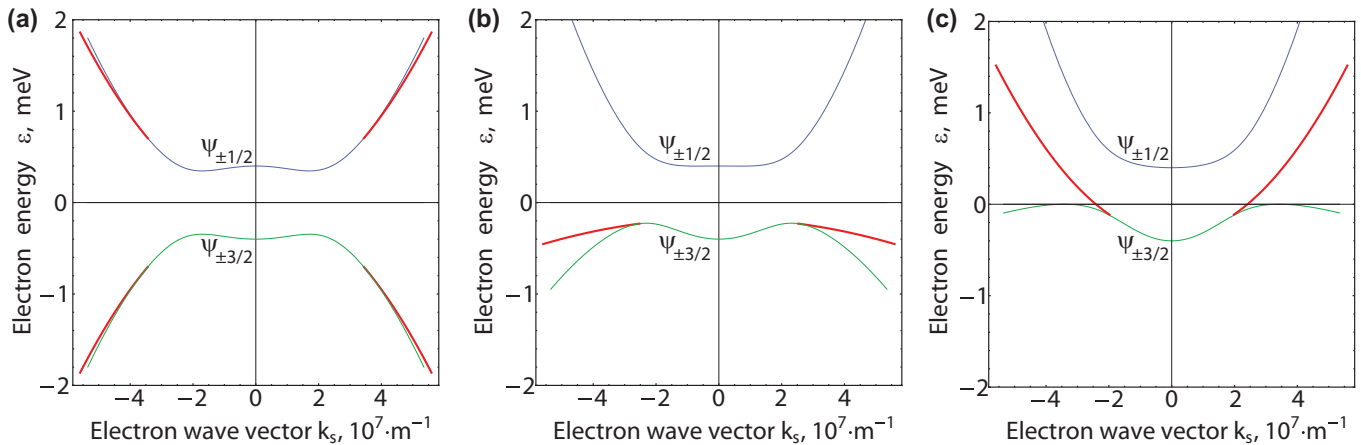}
\caption{Structure of the D'yakonov-Khaetskii surface electronic
states (bold red curves) for the strain-induced gap
$\varepsilon_g=-0.8$~meV and the different values of the Luttinger
parameter $\gamma_1$:(a) $\gamma_1=0$; (b)
$\gamma_1=7.8\,\hbar^2/2m_0$; (c) $\gamma_1=15.6\,\hbar^2/2m_0$
(the case of HgTe). Thin green and blue curves represent the
dispersion of bulk bands which originate from the terms
$\psi_{\pm3/2}$ and $\psi_{\pm1/2}$, respectively.} \label{Fig3}
\end{figure}
%%%

Finally, it should be noted that the experimental methodology
based on the angle resolved photoemission spectroscopy (ARPES)
technique, which is commonly used to study surface electronic
states in various condensed-matter structures~\cite{Chen_2012_1},
can allow to observe the surface states discussed above. It is
important from an experimental viewpoint that the structure of all
considered surface states crucially depends on the bulk Luttinger
parameters, $\gamma_{1,2,3}$. Therefore, they should be chosen
carefully to interpret experimental results adequately in
HgTe-based materials (e.g., solid solutions CdHgTe and MnHgTe),
where the band parameters depend on their stoichiometric
composition.

\section{Conclusion}
We developed the theory describing the structure of various
surface electronic states which appear due to the mixing of the
conduction and valence bands in strained mercury telluride (HgTe).
It predicts the coexistence of surface electronic states of two
different kinds. First of them originate from the bulk inversion
asymmetry of HgTe, have the linear Dirac dispersion, which is
characteristic for topological states, and are localized in very
narrow region of electron wave vectors near the Brillouin zone
center. The surface states of second kind originate from the
D'yakonov-Khaetskii surface states existing in gapless HgTe. Due
to the strain-induced band gap, they are shifted far from the
Brillouin zone center to the region of large electron wave
vectors. Thus, the found surface states of the two kinds exist in
different areas of the Brillouin zone and can be detected
independently. The energy spectrum of the states and their
structure are calculated both analytically and numerically in the
broad range of band parameters.

\begin{acknowledgements}
The work was partially supported by Horizon2020 RISE project
COEXAN, Russian Foundation for Basic Research (project
17-02-00053), Rannis project 163082-051, Ministry of Education and
Science of Russian Federation (projects 3.4573.2017/6.7,
3.2614.2017/4.6, 14.Y26.31.0015), and the Government of the
Russian Federation through the ITMO Fellowship and Professorship
Program. O.V.K. and O.K. thank the University of Iceland for
hospitality.
\end{acknowledgements}

\appendix
\section{The Hamiltonian in the Luttinger-Kohn basis}

The wave functions of the Luttinger-Kohn basis, $\psi_{j_z}$,
corresponding to the different projections of electron angular
momentum on the $z$ axis, $j_z=\pm1/2,\pm3/2$, can be written as
\begin{equation}\label{PsiJ}
\psi_{+3/2}=\frac{1}{\sqrt{2}}(X+iY)\uparrow,\,
\psi_{-1/2}=-\frac{1}{\sqrt{6}}[(X-iY)\uparrow+2Z\downarrow],\,
\psi_{+1/2}=\frac{1}{\sqrt{6}}[(X+iY)\downarrow-2Z\uparrow],\,
\psi_{-3/2}=-\frac{1}{\sqrt{2}}(X-iY)\downarrow,
\end{equation}
where the vertical arrows, $\uparrow$ and $\downarrow$, represent
the spinors corresponding to the $\pm1/2$ spin projections on the
$z$ axis, and $X,Y,Z$ are the Bloch functions in the Brillouin
zone center, which behave like the Cartesian coordinates, $x,y,z$,
under rotation of the coordinate axes~\cite{Bir_Pikus}. In the
basis (\ref{PsiJ}), the angular momentum matrices, $J_{x,y,z}$,
read
\begin{equation}\label{J}
{J}_x=
\begin{bmatrix}
\,\,0\,\, & \,\,0\,\,
&\,\,\frac{\sqrt{3}}{2}\,\,&\,\,0\,\,\\
\,\,0\,\,&\,\,0\,\,&\,\,1\,\,&
\,\,\frac{\sqrt{3}}{2}\,\,\\
\,\,\frac{\sqrt{3}}{2}\,\,&\,\,1\,\,&\,\,0\,\,&
\,\,0\,\,\\
\,\,0\,\,&\,\,\frac{\sqrt{3}}{2}\,\,&\,\,0\,\,&\,\,0\,\,
\end{bmatrix},
\,\, {J}_y=
\begin{bmatrix}
\,\,0\,\, & \,\,0\,\,
&\,\,-i\frac{\sqrt{3}}{2}\,\,&\,\,0\,\,\\
\,\,0\,\,&\,\,0\,\,&\,\,i\,\,&
\,\,-i\frac{\sqrt{3}}{2}\,\,\\
\,\,i\frac{\sqrt{3}}{2}\,\,&\,\,-i\,\,&\,\,0\,\,&
\,\,0\,\,\\
\,\,0\,\,&\,\,i\frac{\sqrt{3}}{2}\,\,&\,\,0\,\,&\,\,0\,\,
\end{bmatrix},
\,\, {J}_z=
\begin{bmatrix}
\,\,\frac{3}{2}\,\, & \,\,0\,\,
&\,\,0\,\,&\,\,0\,\,\\
\,\,0\,\,&\,\,-\frac{1}{2}\,\,&\,\,0\,\,&
\,\,0\,\,\\
\,\,0\,\,&\,\,0\,\,&\,\,\frac{1}{2}\,\,&
\,\,0\,\,\\
\,\,0\,\,&\,\,0\,\,&\,\,0\,\,&\,\,-\frac{3}{2}\,\,
\end{bmatrix}.
\end{equation}
Substituting the matrices (\ref{J}) into Eq.~(\ref{H}), we arrive
at the Hamiltonian (\ref{H}) written in the Luttinger-Kohn basis,
\begin{equation}\label{LHM}
\hat{\cal H}=
\begin{tabular}{|c||c c c c|} \hline
$\psi_{j_z}\backslash \psi_{j_z}$ & $\psi_{+3/2}$ & $\psi_{-1/2}$ & $\psi_{+1/2}$ & $\psi_{-3/2}$ \\
\hline\hline $\psi_{+3/2}$ & $F$ & $I+L$ & $H+M$ & $N$ \\
$\psi_{-1/2}$ & $I^*+L$ & $G$ & $-N$ & $-H+M$ \\
$\psi_{+1/2}$ & $H^*+M^*$ & $-N^*$ & $G$ & $I-L$ \\
$\psi_{-3/2}$ & $N^*$ & $-H^*+M^*$ & $I^*-L$ & $F$ \\
\hline
\end{tabular}\,,
\end{equation}
where
\begin{align}\label{LP}
&F=(\gamma_1+\gamma_2)(k_x^2+k_y^2)+(\gamma_1-2\gamma_2)k_z^2+(a-b)u_{zz}+(a+b/2)(u_{xx}+u_{yy}),\nonumber\\
&G=(\gamma_1-\gamma_2)(k_x^2+k_y^2)+(\gamma_1+2\gamma_2)k_z^2+(a+b)u_{zz}+(a-b/2)(u_{xx}+u_{yy}),\nonumber\\
&I=-\sqrt{3}\gamma_2(k_x^2-k_y^2)+i2\sqrt{3}\gamma_3k_xk_y-(\sqrt{3}b/2)(u_{xx}-u_{yy})+idu_{xy},\,\,M=-(\sqrt{3}\alpha/2)(k_x+ik_y),\nonumber\\
&H=-2\sqrt{3}\gamma_3(k_x-ik_y)k_z-d(u_{xz}-iu_{yz}),\,\,
L=\sqrt{3}\alpha k_z,\,\, N=-(3\alpha/2)(k_x-ik_y).\nonumber
\end{align}
In the particular case of the uniaxial strain along the $z$ axis
and $\gamma_1=0$, the matrix (\ref{LHM}) for $k_x=k_y=0$ turns
into the block-diagonal Hamiltonian (\ref{Hsur}).

\end{document}